\documentclass[12pt]{article}
\usepackage{amsmath}
\usepackage{graphicx,graphics}
\usepackage{enumerate}
\usepackage{natbib}
\usepackage{url} 
\usepackage{amsfonts,amssymb,epsfig,verbatim,bm,latexsym,amsbsy,bbm}
\usepackage{xcolor}

\newcommand{\blind}{0}

\newtheorem{theorem}{Theorem}
\newtheorem{assumption}{Assumption}

\renewcommand{\thesection}{\arabic{section}}
\renewcommand{\theequation}{\arabic{section}.\arabic{equation}}

\newcommand{\R}{I\!\!R}

\DeclareMathOperator*{\plim}{plim}

\allowdisplaybreaks

\begin{document}

\def\spacingset#1{\renewcommand{\baselinestretch}%
{#1}\small\normalsize} \spacingset{1}


\if0\blind
{
  \title{\bf Inference on panel data models with a generalized factor structure}
  \author{Juan M. Rodriguez-Poo\\
    Department of Economics, University of Cantabria\\
    and \\
    Alexandra Soberon\thanks{
    Authors acknowledge financial support from the I+D+i project Ref. PID2019-105986GB-C22 financed by MCIN/AEI/10.13039/501100011033. Further, this work is part of the I+D+i project Ref. TED2021-131763A-I00 financed by MCIN/ AEI/10.13039/501100011033 and by the European Union NextGenerationEU/PRTR.}\hspace{.2cm}\\
    Department of Economics, University of Cantabria\\
        and \\
    Stefan Sperlich \\
    Geneva School of Economics and Management, Université de Genève\\
    }
  \maketitle
} \fi

\if1\blind
{
  \bigskip
  \bigskip
  \bigskip
  \begin{center}
    {\LARGE\bf Inference on panel data models with a generalized factor structure}
\end{center}
  \medskip
} \fi

\bigskip
\begin{abstract}
We consider identification, inference and validation of linear panel data models when both factors and factor loadings are accounted for by a nonparametric function. This general specification encompasses rather popular models such as the two-way fixed effects and the interactive fixed effects ones. By applying a conditional mean independence assumption between unobserved heterogeneity and the covariates, we obtain consistent estimators of the parameters of interest at the optimal rate of convergence, for fixed and large $T$. We also provide a specification test for the modeling assumption based on the methodology of conditional moment tests and nonparametric estimation techniques. Using degenerate and nondegenerate theories of U-statistics we show its convergence and asymptotic distribution under the null, and that it diverges under the alternative at a rate arbitrarily close to $\sqrt{NT}$. Finite sample inference is based on bootstrap. Simulations reveal an excellent performance of our methods and an empirical application is conducted.
\end{abstract}

\noindent%
\vfill

\newpage
\spacingset{1.8} 
\section{Introduction}
\label{sec:intro}

To study the effect of a vector of variables $x$ on output $y$, a very popular model in the panel data literature is
\begin{eqnarray}\label{eqI1a}
y_{it}=x_{it}^{\prime}\beta + m_0\left(\lambda_{i},f_{t}\right) + u_{it},\quad \textrm{for}\quad 1\leq i\leq N,\quad 1\leq t\leq T,
\end{eqnarray}
where $i$ is the cross-section index and $t$ the time series index, $y_{it}$ is a real-valued response variable, $x_{it}$ is a $d_x$-vector of regressors of different typology (i.e., individual-varying, $x_{1i}$, time-varying, $x_{2t}$, and individual-specific covariates, $x_{3it}$), and $\beta$ is a $d_x\times 1$ vector of parameters. The error term comprises two components: $m_0\left(\cdot,\cdot\right)$ is typically a known real-valued function that depends on the unobserved fixed effects, $\lambda_i \in \mathbb{R}^{d_{\lambda}}$, and a vector of unobserved common factors, $f_t \in \mathbb{R}^{d_{f}}$, whereas $u_{it}$ is the idiosyncratic error term. Both vectors, $\lambda_i$ and $f_t$, are allowed to be correlated with the observed regressors $x_{it}$, and $u_{it}$ is supposed to be uncorrelated with $x_{it}$. For instance, $m_0\left(\lambda_{i},f_{t}\right) = \lambda^{\prime}_if_t$ and $m_0\left(\lambda_{i},f_{t}\right) = \lambda_i+f_t$ give the well known interactive fixed effects and the two way fixed effects model, respectively. As in practice it is hard to know the correct specification of $m_0\left(\cdot,\cdot\right)$ one may prefer to think of it as an unknown smooth function.

In fact, model (\ref{eqI1a}) encompasses several panel data models that are of interest in macroeconomics, microeconomics, and finance, among other fields. In macroeconomics, the economic growth of countries is influenced by both individual-specific observable factors ($x_{3it}$) (i.e., physical capital investment, consumption, and population growth), as well as time-varying factors ($x_{2t}$) (i.e., global environmental policies or technological advancements). Furthermore, unobserved common factors ($f_t$) (i.e., global financial crises and fluctuations in world oil prices) exert an impact on all countries through trade and finance linkages with different intensities across countries ($\lambda_i$), as it is noted in \citet{CHUDIK_MOHADDES_PESARAN_RAISSI:2017}. In microeconomics, the individual's wage is determined by a combination of observable characteristics ($x_{3it}$) (i.e., age and working experience) and individual-varying covariates ($x_{1i}$) (i.e., years of education and gender). Furthermore, several unobserved factors, such as innate ability ($\lambda_i$), may be correlated with the observable characteristics, and their causal effects are not constant over time ($f_t$). 

 For $m_0\left(\lambda_{i},f_{t}\right) = \lambda^{\prime}_if_t$, the currently dominant estimation methods for the parameters of interest in (\ref{eqI1a}) are the common correlated effect (CCE) estimator of \citet{PESARAN:2006} and the iterative least squares (ILS) estimator proposed in \citet{BAI:2009} and extended in \citet{MOON_WEIDNER:2015}. The CCE approach proposes to approximate these unobserved factors using linear combinations of cross-sectional averages of the dependent and explanatory variables. Limitations of this method are discussed in \citet{WESTERLUND_URBAIN:2013}. The ILS approach estimates the parameters of interest through a non-convex optimization problem. Recently, least squares estimation with nuclear norm penalization has been proposed to overcome this problem, but the resulting estimator exhibits a fairly slow convergence rate (see \citet{moon2025nuclearnormregularizedestimation} and \citet{beyhum2019squarerootnuclearnormpenalized}). See \citet{CHUDIK:2015}, \citet{KARABIYIK_PALM_URBAIN:2019} and \citet{SARAFIDIS_WANSBEEK:2012} for recent surveys. For $m_0\left(\lambda_{i},f_{t}\right) = \lambda_i+f_t$,  different estimators have been developed; see, for example, Chapter 3 in \citet{HSIAO:2014} and references therein. 
 
On the one side, identification and inference of the parameters of interest in (\ref{eqI1a}) depend crucially on the specification of $m_0\left(\cdot,\cdot\right)$.
On the other side,
when $m_0\left(\cdot,\cdot\right)$ is unknown, identification and 
estimation of $\beta$ in (\ref{eqI1a}) is more cumbersome. 
A popular solution is to introduce a conditional mean independence restriction between $(\lambda_i,f_t)$ and $x_{it}\in \mathbb{R}^{d_x}$. This approach is in spirit related to the nonparametric control function approach (see, for instance, \citet{NEWEY_POWELL_VELLA:1999} in a different context), and even more to the extended Mundlak device for mixed-effects models (see \citep{LombardiaEtal:2012}). Under conditional mean independence, model (\ref{eqI1a}) takes the form of a partial linear model \citep{HARDLE_LIANG_GAO:2000} with some advantages in terms of identification, inference and validation. 

In this way, we can propose a consistent specification test that checks whether this assumption is credible. The test relies on ideas of \citet{ZHENG:1996}, \citet{SU_JIN_ZHANG:2015}, and \citet{CAI_FANG_XU:2020}, among others, combining the methodology of conditional moments tests and nonparametric estimation techniques. Using degenerate and nondegenerate theories of U-statistics we can show the asymptotic distribution of our test under the null, and that it diverges under the alternative at a rate of order $\sqrt{NTh^{d/2}}$,
where $h$ is a smoothing parameter and $d\equiv d_x+d_w$. 
If the conditional mean independence assumption is fulfilled, by using semiparametric estimation techniques for partial linear models, such as the profile least squares approach proposed in \citet{FAN_HUANG:2005}, one can obtain relatively straight-forward but efficient estimators for $\beta$.

In sum, the introduction of a conditional mean independence restriction between $(\lambda_i,f_t)$ and $x_{it}$ converts model (\ref{eqI1a}) to a simple partially linear model for panel data. Such semiparametric specification has important advantages, mainly: i) We obtain consistent estimators of the parameters of interest, $\beta$, with optimal rates of convergence; ii) We can identify the parameters of interest even in the presence of individual-varying or time-varying covariates since the method is not based on quasi-differencing techniques; iii) Our asymptotic results are obtained under two different scenarios: $N,T \rightarrow \infty$ and for fixed $T$, $N\rightarrow \infty$; and iv) we can derive a specification test for the conditional mean independence assumption.

We employ our methods to assess the effect of the European Union Emissions Trading System (EU-ETS) on the economic development of EU countries. This topic is of great interest since this policy is the cornerstone of the EU's strategy to decarbonize the economy, but there is a certain reluctance due to its potentially adverse economic impact. We find that the postulates of the Environmental Kuznets Curve are corroborated, together with the significant effect of population and energy intensity on environmental degradation. Moreover, our results show the inability of the typically imposed functional forms for fixed effects interactions to capture the effect of unobserved common factors in a proper way, which results in counter-intuitive, misleading conclusions.

Section \ref{secII} proposes our modeling approach with estimation procedures, and establishes the asymptotic distributions of our estimators. Section \ref{secIII} presents a test to check the conditional independence assumption, and introduces a bootstrap method to perform inference in finite samples. Section \ref{secIV} studies the finite sample performance of the proposed estimator and test statistic via several Monte Carlo experiments. Section \ref{secV} studies the effects of the EU ETS. Section \ref{secVI} concludes. The proofs of the main theorems are shown in Appendix A. The remaining theoretical lemmas with proofs with additional simulation studies are provided in the Supplementary Material.


\section{Estimation procedure and asymptotic properties}\label{secII}
\setcounter{equation}{0}

\subsection{Modeling and estimator}\label{secIIa}

Consider model (\ref{eqI1a}) with $m_0$ being an unknown smooth function. Following the discussion in the Introduction, for identification of $\beta$ we introduce a conditional independence assumption along a set of observed variables
$w_{it} \in \mathbb{R}^{d_w}$, such that
\begin{assumption}\label{asumAA1}
For $1\leq i\leq N$ and $1\leq t\leq T$, $E\left(\left.u_{it}\right|x_{it},w_{it},\mathbb{D}\right)=0$, where $\mathbb{D} \equiv \sigma\left(\{\lambda_i\},\{f_t\}\right)$ is the minimal $\sigma$-field generated from $\{f_t\}_{t=\cdots,-1,0,1,\cdots}$ and $\{\lambda_i\}^N_{i=1}$. 
\end{assumption}
This condition implies exogeneity of $u_{it}$ with respect to the pair $(x_{it},w_{it})$, i.e., $E\left(\left.u_{it}\right|x_{it},w_{it}\right)=0$ and that both factors and factor loadings 
are exogenous to the error term $u_{it}$, i.e., $E\left(\left.u_{it}\right|
\mathbb{D}\right)=0$. Next, we introduce an assumption about the relationship between factors, factor loadings, and covariates 
\begin{assumption}\label{asumAA2}
$E\left[m_0\left(\lambda_i,f_t\right)\right] < \infty$, $E\left[\left.m_0\left(\lambda_i,f_t\right)\right|x_{it} \right] < \infty$,
$E\left[\left.m_0\left(\lambda_i,f_t\right)\right|x_{it},w_{it} \right] = E\left[\left.m_0\left(\lambda_i,f_t\right)\right|w_{it} \right]$, and $E\left[\left.m_0\left(\lambda_i,f_t\right)\right|w_{it} \right] = g\left(w_{it}\right)$,
 for $i=1,\cdots,N$ and $t=1,\cdots,T$, where $g(\cdot)$ is a real-valued function.  
\end{assumption}
This implies that conditionally on $w_{it}$, $(\lambda_i, f_t)$ is mean independent of $x_{it}$. As said, it is a frequently used strategy for the identification of $\beta$, as in the case of control variables, control functions, or the Mundlak device. In practice, popular examples for (the elements of) $w_{it}$ are $\bar{x}_i$, $\bar{x}_t$ or their product. Depending on the context, other examples can be the inflation rate or, as in our application, the stock of debt liabilities.  

Taking conditional expectations on both sides of (\ref{eqI1a}) with respect to $(x_{it},w_{it})$, and applying Assumptions \ref{asumAA1} and \ref{asumAA2}, equation (\ref{eqI1a}) becomes
\begin{eqnarray}\label{eqI1b}
y_{it}=x_{it}^{\prime}\beta+ g\left(w_{it}\right) + \epsilon_{it},\quad \textrm{for}\quad 1\leq i\leq N,\quad 1\leq t\leq T,
\end{eqnarray}
where
$\epsilon_{it} = y_{it} - E\left(\left.y_{it}\right|x_{it},w_{it}\right) = u_{it} + \eta_{it}$ 
and
$\eta_{it} = m_0\left(\lambda_i,f_t\right)-g\left(w_{it}\right)$. 

Here, $x_{it}$ and $w_{it}$ represent different features: $x_{it}$ contains the observed regressors of interest, and $w_{it}$ observed or constructed covariates that may facilitate the control for unobserved factors.

To obtain a consistent estimator for $\beta$, we propose a profile least-squares estimation procedure following ideas of \citet{FAN_HUANG:2005} 
since among the known estimation procedures for partial linear models, it turned out to adapt quite well to our problem. 
More precisely, for a given value of $\beta$, we can propose a local linear least squares kernel estimator of the nonparametric function $g(\cdot)$ at a point $w\in \mathbb{R}^{d_w}$, $\widehat{a}$, as the solution of 
\begin{eqnarray}\label{eqII2}
\arg\min_{a,b}\sum_{i=1}^N\sum_{t=1}^T\left[(y_{it}-x_{it}^{\prime}\beta)-a-(w_{it}-w)^{\prime}b\right]^2K_{h_w}(w_{it}-w_0),\quad
\end{eqnarray}
where $K_{h_w}(\cdot)$ is a kernel function. For multivariate $w$ we use a product kernel $K_{h_w}(u)=\prod_{l=1}^{d_w}k_{h_w}(u_{l})$ with $u=(u_1,\ldots,u_{d_w})^{\prime}$
and $k_{h_w}(w_{it}-w)=h_w^{-1}k((w_{it}-w)/h_w)$.

For any given $\beta$, the minimizer of (\ref{eqII2})
leads to the infeasible estimator for $g(\cdot)$
\begin{eqnarray}\label{eqII3}
\widetilde{g}(w; h_w)&=&(1,0_{d_w})[\mathcal{W}^{\prime}K_w\mathcal{W}]^{-1}\mathcal{W}^{\prime}K_w[Y-X\beta],
\end{eqnarray}
where $Y=(y_{11},\ldots,y_{NT})$ is a $NT\times 1$ vector, $\mathcal{W}\in\R^{(1+d_w)}$ whose $it$-th element is $[1,(w_{it}-w_0)^{\prime}]$, and $X=(x_{11},\ldots,x_{NT})$ is a $NT\times d_x$ matrix. In addition, $K_{w}=diag\{K_{h_w}(w_{11}-w),\ldots,K_{h_w}(w_{NT}-w)\}$ is a $NT\times NT$ diagonal matrix and $0_{d_w}$ a $d_w$-vector of zeros.

Writing $\widetilde{g}(w; h_w)$ in terms of the kernel projection $S\in \R^{NT \times NT} $ defined below, and plugging the resulting expression into model (\ref{eqI1b}), we obtain a transformed regression model that can be written in vectorial form as
\begin{eqnarray}\label{eqII4}
	\widetilde{Y}=\widetilde{X}\beta+\epsilon^*,
\end{eqnarray}
where $\widetilde{Y}=(I_{NT}-S)Y$ and $\epsilon^*$ are $NT\times 1$ vectors, whereas
$\widetilde{X}=(I_{NT}-S)X$ with $X$ and $\widetilde{X}$ being $NT\times d_x$ matrices. Also, it is not hard to see that $\epsilon_{it}^*=\epsilon_{it}+\widetilde{g}(w;h_w)-g(w_{it})$ and
\begin{eqnarray*}
S=\left[
\begin{array}{c}
(1,0_{d_w})\left[\mathcal{W}^{\prime}K_{w_{11}}\mathcal{W}\right]^{-1}\mathcal{W}^{\prime}K_{w_{11}}\\
\vdots\\
(1,0_{d_w})\left[\mathcal{W}^{\prime}K_{w_{NT}}\mathcal{W}\right]^{-1}\mathcal{W}^{\prime}K_{w_{NT}}
\end{array}\right]  \ .
\end{eqnarray*}

Then, from (\ref{eqII4}), the least squares estimator proposed for $\beta$ is
\begin{eqnarray}\label{eqII5}
\widehat{\beta}=[X^{\prime}(I_{NT}-S)^{\prime}(I_{NT}-S)X]^{-1}X^{\prime}(I_{NT}-S)^{\prime}(I_{NT}-S)Y  \ .
\end{eqnarray}

Finally, while $\widetilde{g}(w;h_w)$ is an infeasible estimator because it depends on the unknown parameter $\beta$, plugging (\ref{eqII5}) in (\ref{eqII3}) gives a feasible estimator of the form
\begin{eqnarray}\label{eqII6}
\widehat{g}(w; h_w)&=&(1,0_{d_w})[\mathcal{W}^{\prime}K_w\mathcal{W}]^{-1}\mathcal{W}^{\prime}K_w[Y-X\widehat{\beta}]   \ .
\end{eqnarray}


\subsection{Asymptotic properties}\label{secIIc}

Let us consider the two standard scenarios: (i) $N\rightarrow\infty$ and $T$ fixed; (ii) $N\rightarrow\infty$ and $T\rightarrow\infty$. For the latter case, we first recall the definition of a strongly mixing sequence. 

Let $\{\zeta_t\}$ be a strictly stationary process and $\mathcal{F}_{t'}^t$ denotes a $\sigma$-algebra of events generated by the random variables $(\zeta_{t'},\ldots,\zeta_t)$ for $t'\leq t$. Following \citet{ROSENBLATT:1956}, a process is said to be strongly mixing or $\alpha$-mixing if
\[
\alpha(\tau)=\sup_{t'\in\mathcal{N}}\{|P(A\cap B)-P(A)P(B)|:A\in\mathcal{F}_{-\alpha}^{t'},B\in\mathcal{F}_{t'+\tau}^{-\infty}\}\rightarrow0,\quad\textrm{as}
\quad T\rightarrow\infty  \ .
\]

For the sake of presentation, we introduce now the following notation: $B_x(w_0)=E[x_{it}|w_{it}=w_0]$, $\Phi_{\epsilon}(\chi_{it},\chi_{it'})=E(\epsilon_{it}\epsilon_{it'}|\chi_{it},\chi_{it'})$, and $\Omega_x=E[\{x_{it}-B_x(w_{it})\}\{x_{it}-B_x(w_{it})\}^{\prime}]$, for $\chi_{it}=(x_{it},w_{it})$. For the univariate kernel $k(\cdot)$, we denote $\mu_2=\int u^2k(u)du$, $\nu_0=\int k^2(u)du$, and $\nu_2=\int u^2k^2(u)du$, where $\mu_2$, $\nu_0$, and $\nu_2$ are scalars different from zero. Further, for a real matrix \textbf{A}, let $\|A\|=tr^{1/2}(A^{\prime}A)$. For a vector $\textbf{v}$, let $\|\textbf{v}\|$ denote its Euclidean norm. Also, let $D_g(w)$ and $\mathcal{H}_g(w)$ be the $d_w\times 1$ first-order derivative vector and the $d_w\times d_w$ Hessian matrix of $g(\cdot)$ with respect to $w$, respectively. Similarly, let $D_{\rho}(w)$ be the $d_w\times1$ first-order derivative vector of $\rho(w)$. 

Then, to show $\sqrt{N}$-consistency of $\widehat \beta$ as $N$ tends to infinity and $T$ is fixed, we suppose

\begin{assumption}\label{asumA1}
The univariate kernel function $k(\cdot)$ is a symmetric, non-negative, and bounded probability density function with compact support which satisfies the Lipschitz condition, $\int k(v)dv=1$, and $\int vk(v)dv=0$. 
\end{assumption}
\begin{assumption}\label{asumA2}
Let $\{(x_{it},w_{it},u_{it},\eta_{it}),i\geq1,t\geq 1\}$ be a set of independent and identically distributed ($i.i.d.$) random variables across the $i$ index for each fixed $t$. Also, for fixed $i$, the process $(x_{it},w_{it},u_{it},\eta_{it})$ is strictly stationary and ergodic over $t$. Furthermore, $u_{it}$ and $\eta_{it}$ are independent.
\end{assumption}

\begin{assumption}\label{asumA3}
Let $\varrho_{it} =(u_{it}, \eta_{it})$ and $\chi_{it} = (w_{it},x_{it})$. 
$\exists$ 
$\gamma>0$, such that  $E(\left|\varrho_{it}|^{4+\gamma}\right|\chi_{it})<\infty$,
$\forall$ 
$i$ and $t$. Also, $E|\ddot{x}_{\kappa,it}\varrho_{it}|^{2(1+\gamma)}<\infty$, where $\ddot{x}_{it}=x_{it}-B_x(w_{it})$ and $\kappa=1,\ldots,d_x$.
\end{assumption}
\begin{assumption}\label{asumA4}
\hfill
\vspace{-0.4cm}
\begin{itemize}
\item[a)] $w_{it}$ has compact support, $\mathbb{W}$, and its density function, $\rho_w(\cdot)$, is Lipschitz continuous and bounded away from zero, i.e., $\rho_w(w)>0$ for each $w \in \mathbb{W}$.
\item[b)] The functions $\rho_w(\cdot)$ and $g(\cdot)$ have bounded continuous derivatives up to the second order. Also, $B_x(w)$ has continuous second derivatives with respect to $w$ for $w\in\mathbb{W}$.
\end{itemize}
\end{assumption}
\begin{assumption}\label{asumA5}
The matrix $\Omega_x$ is nonsingular. There exists $\gamma^{\prime}>2$ such that 
\[
\sup_{w\in\mathbb{W}}\{E[\|x_{it}\|^{2\gamma'}|w_{it}=w]\}<\infty, 
\]
and $N^{(2\varepsilon-1)}h_w\rightarrow\infty$ for some $\varepsilon<1-(2+\gamma')^{-1}$.
\end{assumption}
\begin{assumption}\label{asumA6}
$h_w\rightarrow0$ and $NTh_w^{d_w}\rightarrow\infty$ as $N\rightarrow\infty$ and $T$ is fixed.
\end{assumption}

These assumptions are fairly standard for local linear estimation in panel data models with $T$ fix. In particular, Assumptions \ref{asumA1}, \ref{asumA4} a), and \ref{asumA5} are sufficient to show the uniform convergence of a kernel-type regression estimator (see \citet{MACK_SILVERMAN:1982}). 

We can also show the $\sqrt{NT}$-consistency of $\widehat \beta$ when both $N$ and $T$ tend to infinity. In order to do so, we replace Assumptions \ref{asumA2}, \ref{asumA3}, and \ref{asumA6} by the following ones:

\begin{assumption}\label{asumB1}
Let $\{(x_{it},w_{it},u_{it},\eta_{it}),i\geq1,t\geq 1\}$ be a set of independent and identically distributed ($i.i.d.$) random variables across the $i$ index for each fixed $t$. For each fixed $i$, the process $(x_{it},w_{it},u_{it},\eta_{it})$ is $\alpha$-mixing with $\sum_{\tau=1}^{T-1}[\alpha(\tau)]^{\gamma/(1+\gamma)}<\infty$ for some $\gamma>0$. Also, the mixing coefficient satisfies the condition $\alpha(\tau)=O(\tau^{-\vartheta})$, where $\vartheta=(2+\gamma)(1+\gamma)/\gamma$. Furthermore, $u_{it}$ and $\eta_{it}$ are independent.
\end{assumption}

\begin{assumption}\label{asumB2}
For $\ddot{x}_{it}=x_{it}-B_x(w_{it})$, $\max_{t\leq T}E|\ddot{x}_{\kappa,it}\varrho_{it}|^{2(1+\gamma)}<\infty$, for $\varrho_{it} = (u_{it}, \eta_{it})$, where $\kappa=1,\ldots,d_x$ and $\gamma>0$.
\end{assumption}

\begin{assumption}\label{asumB3}
As $(N,T)\rightarrow\infty$, $h_w\rightarrow0$ and $NTh_w^{d_w}\rightarrow\infty$.
\end{assumption}

To obtain the asymptotic distribution of the nonparametric estimator (\ref{eqII6}) we add

\begin{assumption}\label{asumD1}
As $(N,T)\rightarrow \infty$, $NT^{(\vartheta+1)/\vartheta}h_w^{d_w(2+\gamma)/(1+\gamma)}\rightarrow\infty$ for $\gamma>0$ and $\vartheta=(2+\gamma)(1+\gamma)/\gamma$.
\end{assumption}

The following theorems collect the main asymptotic properties of the estimator of the parameter of interest, $\beta$. Detailed proofs of them can be found in the Appendix A.
\begin{theorem}\label{theo1}
\hfill
\begin{itemize}
\item[(i)] Under Assumptions \ref{asumA1}-\ref{asumA6}, if $N\rightarrow\infty$ and $T$ is fixed, we have
\begin{eqnarray*}
\sqrt{NT}\left(
\widehat{\beta}-\beta\right)\overset{d}{\to}\mathcal{N}(0,\Omega_x^{-1}\Sigma_{1,\epsilon}\Omega_x^{-1}),
\end{eqnarray*}
where $\Sigma_{1,\epsilon}=\frac{1}{T}\sum_{t=1}^T\sum_{t'=1}^T\Phi_{\epsilon}
(\chi_{it},\chi_{it'})\Omega_x$.
\item[(ii)] Under Assumptions \ref{asumA1}, \ref{asumA4}-\ref{asumA5}, and \ref{asumB1}-\ref{asumB3}, if $N$ and $T$ tend to infinity, we have
 \begin{eqnarray*}
\sqrt{NT}\left(
\widehat{\beta}-\beta\right)\overset{d}{\to}\mathcal{N}(0,\Omega_x^{-1}\Sigma_{2,\epsilon}\Omega_x^{-1}),
\end{eqnarray*}
where $\Sigma_{2,\epsilon}=\displaystyle\plim_{T\rightarrow\infty}\frac{1}{T}\sum_{t=1}^T\sum_{t'=1}^T
\Phi_{\epsilon}(\chi_{it},\chi_{it'})\Omega_x$.
\end{itemize}
\end{theorem}

\begin{theorem}\label{theo2}
\hfill
\begin{itemize}
\item[(i)] Under Assumptions \ref{asumA1}-\ref{asumA6}, if $\sqrt{NTh_w^{(d_w+8)}}=O(1)$, as $N\rightarrow\infty$ and $T$ is fixed, 
\begin{eqnarray*}
\sqrt{NTh_w^{d_w}}\left(\widehat{g}(w;h_w)-g(w)-\frac{h_w^2\mu_2^{d_w}}{2}tr\{
\mathcal{H}_{g}(w)\}\right)\overset{d}{\to}\mathcal{N}\left(0,
\frac{\nu_0^{d_w}\Psi_{1,\epsilon}}{\rho_w(w)}\right)
\end{eqnarray*}
where $\Psi_{1,\epsilon}=\frac{1}{T}\sum_{t=1}^T\sum_{t'=1}^TE(\epsilon_{it}
\epsilon_{it'}|w_{it},w_{it'})$.
\item[(ii)] Under Assumptions \ref{asumA1}, \ref{asumA4}-\ref{asumA5}, \ref{asumB1}, and \ref{asumB3}-\ref{asumD1}, if $\sqrt{NTh_w^{(d_w+8)}}=O(1)$, as $N$ and $T$ tend to infinity, we have
\begin{eqnarray*}
\sqrt{NTh_w^{d_w}}\left(\widehat{g}(w;h_w)-g(w)-\frac{h_w^2\mu_2^{d_w}}{2}tr\{
\mathcal{H}_{g}(w)\}\right)\overset{d}{\to}\mathcal{N}\left(0,
\frac{\nu_0^{d_w}\Psi_{2,\epsilon}}{\rho_w(w)}\right)
\end{eqnarray*}
where $\Psi_{2,\epsilon}=\displaystyle\plim_{T\rightarrow\infty}\frac{1}{T}\sum_{t=1}^T\sum_{t'=1}^TE(\epsilon_{it}
\epsilon_{it'}|w_{it},w_{it'})$.
\end{itemize}
\end{theorem}


\section{Validation, Bootstrap and Extensions}\label{secIII}
\setcounter{equation}{0}

Above we have shown that Assumptions \ref{asumAA1} and \ref{asumAA2} are sufficient to identify the parameters of interest in model (\ref{eqI1a}).
Since Assumption \ref{asumAA1} is the basic one, it would be desirable to have a testing device that checks the credibility of the crucial second one. 

\subsection{A Nonparametric Specification Test}

Testing Assumption \ref{asumAA2} is equivalent to testing the  hypothesis 
\begin{equation}\label{h0}
H_0: \  E\left(\left. \epsilon_{it} \right|\chi_{it}\right)  = 0 \quad \textrm{a.s}
   \qquad  \mbox{ against alternative }  \qquad 
H_1: \  E\left(\left. \epsilon_{it} \right|\chi_{it}\right)  \ne 0 \quad \textrm{a.s}.
\end{equation}

This is because by Assumption \ref{asumAA1}, $E\left(\left.u_{it}\right|\chi_{it}\right)=0$, but if Assumption \ref{asumAA2} is not fulfilled, $E\left(\left.\eta_{it}\right|\chi_{it}\right) \ne 0$. Thus, the alternative encompasses all the possible departures from the null model. Let $\rho_{\chi}\left(\chi_{it}\right)$ be the probability density function of $\chi \in \mathbb{R}^d$, where $d=d_x+d_w$. Under $H_0$, since $E\left(\left. \epsilon_{it} \right|\chi_{it}\right)  = 0$, we have
\begin{equation}
E\left[\epsilon_{it}E\left(\left.\epsilon_{it}\right|\chi_{it}\right)\rho_{\chi}\left(\chi_{it}\right)\right] = 0,
\end{equation}
whereas under $H_1$, since $E\left(\left.\epsilon_{it}\right|\chi_{it}\right) =
E\left[\left.m_0\left(\lambda_i,f_t\right)\right|\chi_{it}\right] - g(w_{it}) \ne 0$, we have
\begin{eqnarray}
\nonumber  E\left[\epsilon_{it}E\left(\left.\epsilon_{it}\right|\chi_{it}\right)\rho\left(\chi_{it}\right)\right] & = & E\left\{\left[E\left(\left.\epsilon_{it}\right|\chi_{it}\right)\right]^2\rho\left(\chi_{it}\right)\right\}  \\
\nonumber  & = & E\left\{\left[ E\left(\left.m_0\left(\lambda_i,f_t\right)\right|\chi_{it}\right) - g(w_{it})\right]^2\rho_{\chi}\left(\chi_{it}\right)\right\}  \ > \ 0 \ .
\end{eqnarray}

A sample analogue of $E\left[\epsilon_{it}E\left(\left.\epsilon_{it}\right|\chi_{it}\right)\rho_{\chi}\left(\chi_{it}\right)\right]$ is
\begin{equation}\label{eqvn}
    V_{NT} = \frac{1}{T^2N^2h^d}\sum^N_{i=1}\sum_{j\neq i}^N \sum^T_{t=1}\sum^T_{s=1} K_{it,js} \widehat{\epsilon}_{it}\widehat{\epsilon}_{js}  \ ,
\end{equation}
where again we use a product kernel, namely
\[
K_{it,js} = K\left(\frac{\chi_{it}-\chi_{js}}{h}\right); \quad K(v) = \prod^m_{l=1}k(v_l)
\]
with $h$ being the bandwidth, and $\widehat{\epsilon}_{it}$ the residuals, i.e., $\widehat{\epsilon}_{it} = y_{it} - x^{\prime}_{it}\widehat{\beta}- \widehat{g}(w_{it};h_w)$.

To derive asymptotic properties of $V_{NT}$, the following assumptions are used.
\begin{assumption}\label{aa1}
Let $\delta_1 > 4$,  $\left(E\left|\varrho_{it}\right|^{\delta_1}\right)^{1/\delta_1} < \infty$. Given $i$, $\alpha_{\varrho}(\tau)$ is the $\alpha$-mixing coefficient of $\left\{\varrho_{it}: t\ge 1\right\}$ such that $\sum^{\infty}_{\tau=1}\tau\alpha_{\varrho}\left(\tau\right)^{(\delta_1-4)/\delta_1}<\infty$.
\end{assumption}

Then the asymptotic distribution under the null and the power of our test are given by the following theorems.
\begin{theorem}\label{theo5}
    Under Assumptions \ref{asumA1}, \ref{asumA4}-\ref{asumA5}, \ref{asumB1}, \ref{asumB3}-\ref{asumD1} and \ref{aa1}, if $h \rightarrow 0$, $NTh^d\rightarrow \infty$ and
    $\left(NT\right)^{d/d+4}h^{d/2}\left(\mathrm{log}NT\right)^{4/d+4}\rightarrow 0$, as $N$ and $T$ tend to infinity, then under the null hypothesis (\ref{h0}), we have
    \begin{equation}\label{eqad}
        NTh^{d/2}V_{NT}/\widehat\upsilon_0 \overset{d}{\to}\mathcal{N} \left(0,1\right),
    \end{equation}
    where
\begin{equation}\label{eqavv}
\widehat\upsilon^2_0 = \frac{2}{N^2T^2h^d}\sum^N_{i=1}\sum_{j\neq i}^N \sum^T_{t=1}\sum^T_{s=1}K^2_{it,js}\widehat{\epsilon}^2_{it}\widehat{\epsilon}^2_{js}
\end{equation}
is a consistent estimator of the asymptotic variance of $NTh^{d/2}V_{NT}$, i.e.,
  \begin{equation}\label{eqav}
      \upsilon^2_0 = \lim_{T\rightarrow\infty}\frac{2}{T^2h^{d}}\sum^T_{t=1}\sum^T_{s=1}E\left(K^2_{1t,2s}\epsilon^2_{1t}\epsilon^2_{2s}\right).
  \end{equation}
\end{theorem}

\begin{theorem}\label{theo6}
    Under Assumptions \ref{asumA1}, \ref{asumA4}-\ref{asumA5}, \ref{asumB1}, \ref{asumB3}-\ref{asumD1} and \ref{aa1}, if $h \rightarrow 0$, $NTh^d\rightarrow \infty$, and
    $\left(NT\right)^{d/d+4}h^{d/2}\left(\mathrm{log}NT\right)^{4/d+4}\rightarrow 0$, as $N$ and $T$ tend to infinity, then under the alternative $H_1$, we have
    \begin{equation}\label{eqad2}
       \textrm{Pr}\left\{NTh^{d/2}V_{NT}/\widehat{\upsilon}_0 \ge \delta_{NT}\right\} \rightarrow 1,
    \end{equation}
where $\delta_{NT}$ is any non-stochastic positive sequence such that $\delta_{NT} = o\left(NTh^{d/2}\right)$.
\end{theorem}

While on a first glimpse this seems to be an extremely strong result, indicating one could fully validate by a data-driven specification test the key identification condition, some caution is recommended. 
In fact, it is known that this type of identification conditions cannot be fully tested; instead one resorts to tests that give or rest credibility to the condition in question. In our case, for instance, it is not hard to see that under $H_1$, the power of our test hinges on the parts of $\eta$ that are either nonlinear in $x$ or reflect interactions between $x$ and $w$.  
Specifically, the stronger these are emphasized, the easier we reject, else we may not. This tells us how and when we can detect violations of Assumption \ref{asumAA2}.

\subsection{Bootstrap and Bandwidth}
The asymptotic results for our estimators and test statistic are very helpful in showing consistency and convergence rate, and to better understand its statistical behavior. Nevertheless, it is less helpful for doing further inference in finite samples because estimates of those first-order asymptotic bias and variance are quite poor approximates in practice. A commonly employed remedy are resampling methods. This suggests using the so-called wild bootstrap, see \cite{MAMMEN:1992}. Apart from being consistent for statistics based on local estimators like ours, it accounts for potential heteroscedasticity and allows us to accommodate nonparametric dependence structures in the error term. Recall that we only require independence for given $t$ but not between errors referring to the same individual.

First calculate $\widehat{\beta}$ and $\widehat{g}(w_{it})$ for all $w_{i,t}$ of the sample, obtaining residuals $\widehat{\epsilon}_{it}$.
Next, for the given sample of regressors $\{ \chi_{it} = (x'_{it},w'_{it})' \}_{i=1,t=1}^{N,T}$, generate bootstrap output samples $\{ y_{it}^{*,b} \}_{i=1,t=1}^{N,T}$ by drawing $\vartheta_i^b \stackrel{i.i.d.}{\sim} N(0,1)$ for $b=1,\ldots,B$ and setting
\begin{eqnarray}\label{eqboot}
 y_{it}^{*,b} &:=&  x'_{it}\widehat{\beta} + \widehat{g}(w_{it}) +
 \widehat{\epsilon}_{it} \vartheta_i^b \ .
\end{eqnarray}
It is easy to see that $E[\widehat{\epsilon}_{it} \vartheta_i^b] =0$, $Cov[\widehat{\epsilon}_{it} \vartheta_i^b, \widehat{\epsilon}_{js} \vartheta_j^b] = Cov[\widehat{\epsilon}_{it},\widehat{\epsilon}_{js}]$, for all $i$ and $j=1,\ldots,N$, $t$ and $s=1,\ldots,T$, i.e., the first two moments and dependence is maintained in the bootstrap. Third, calculate $\widehat{\beta}^{*,b}$ and $\widehat{g}^{*,b}(w_{it})$ for all bootstrap samples $b=1,\ldots,B$. Their sample means and variances give the bootstrap estimates of bias and variance of our estimators.

In the same way one can calculate bootstrap p-values for the test by comparing $V_{NT}/\widehat{\upsilon}_0$ with its bootstrap analogs $V_{NT}^{*,b}/\widehat{\upsilon}_0^{*,b}$. Note that the bootstrap samples are generated under the null hypothesis such that the bootstrap analogs follow approximately the finite sample distribution of $V_{NT}/\widehat{\upsilon}_0$ under $H_0$. Consequently, the p-value is the proportion of bootstrap statistics bigger than the original one.

For bootstrap tests, it is often recommended to employ in (\ref{eqboot})
residuals $\widehat{\epsilon}_{it}$ obtained under the alternative. On the one hand, it is supposed this increased the power of the test, but in practice, this tends to produce over-rejection  \citep{SPERLICH:2014}. In our case, it is not even clear what a good predictor of residuals under the alternative could be. Therefore we stick to the former, more common practice of using the residuals obtained under $H_0$.

Regarding the bandwidth choice for estimation, it is to be noted that from a theoretical point of view, there does not exist a 'generally optimal' bandwidth since what is 'optimal' depends on the specific objective. More precisely, for estimating $\widehat{\beta}$ a different bandwidth is optimal than for estimating $\widehat{g}(w_{it})$, and both are different from the optimal testing bandwidths, not to mention the bandwidths optimal for generating the bootstrap samples \citep{SPERLICH:2014}. If the null hypothesis is non- or semiparametric, as it is in our case, then it can become quite tedious to calibrate the test along all those different bandwidth choices as it is noted in \citet{RODRIGUEZ-POO_SPERLICH_VIEU:2015}. In practice, we should be pragmatic and prefer a method that delivers reasonable estimates and provides a well-functioning test procedure. Consequently, we abstain here from the search for optimal bandwidths but propose a simple and computationally attractive solution. Then, it is helpful to recall that typically, the main interest is in $\beta$, not in function $g(\cdot)$. With this focus in mind, it turns out (see our simulations) that Silverman's rule-of-thumb, though invented for density estimation, is very useful in this context. When using for $k(\cdot )$ the Epanechnikov kernel, then the proposed bandwidth is $h_{w}=2.345\widehat{\sigma}_w(NT)^{-1/5}$, with $\widehat{\sigma}_w$ being the sample standard deviation of $w_{it}$. For more sophisticated bandwidth selection procedures, e.g., to estimate $g(\cdot )$ in an optimal way, we refer to the review of \cite{KOEHLER_SCHINDLER_SPERLICH:2014}.


\section{Monte Carlo simulations} \label{secIV}

The aim is to illustrate the performance of our proposed estimators and test in finite samples using simulated data. We further compare our estimator with alternatives proposed in related panel data literature (i.e., the CCE estimator of \citet{PESARAN:2006} and the principal component approach (PCA) of \citet{BAI:2009}). Hence, we employ the following data-generating processes (DGP)
\begin{eqnarray}
y_{it}&=&x_{it}^{\prime}\beta_1+z_t^{\prime}\beta_2+m_0\left(\lambda_i,f_t\right)+u_{it},\label{eqIV1}
\end{eqnarray}
for $i=1,\ldots,N$, $t=1,\ldots,T$. Note that $m_0\left(\lambda_i,f_t\right)=-w_{it}^2+2w_{it}+\xi_{m,it}$, where $\xi_{m,it}\sim IIDN(0,1)$, and each experiment has been replicated $1000$ times for $N=\{20,30,50,100\}$ and $T$ to be either $\{20,30,40,50,100\}$. In Section S1.1 of the Supplement, we specify how to generate the regressors $(x_{it},z_t)$, unobserved factors $(f_t)$, factor loadings $(\lambda_i)$, and individual-specific errors $(u_{it})$.

Although different choices of kernels and bandwidths would be feasible, we use the Epanechnikov kernel $k(u)=0.75(1-u^2)\mathbbm{1}\{|u|\leq1\}$ together with bandwidth $h_w=I_p h_{w}$, where $h_{w}$ is the Silverman's rule-of-thumb bandwidth as proposed above.
For evaluation of the performance of our estimators, we use the bias and the root mean squared errors (RMSE) for the slope parameters, while the median of the RMSE is computed for the regression functions. For model (\ref{eqIV1}) we collect the results for $\beta_1$ in Tables \ref{tabIV1}-\ref{tabIV2}. The results for $\beta_2$ are very similar and therefore not reported here.

\begin{table}[!ht]  		
  \centering
  \caption{Small sample properties of the parametric estimators in (\ref{eqII5})}\label{tabIV1}
  \resizebox{12cm}{!}{
  \begin{tabular}{lrrrrrrrrr}
  \hline
&\multicolumn{4}{c}{Bias (x100)} &&\multicolumn{4}{c}{RMSE}\\
\hline
N/T   &\multicolumn{1}{c}{20} &\multicolumn{1}{c}{30} &\multicolumn{1}{c}{50} &\multicolumn{1}{c}{100}  &&\multicolumn{1}{c}{20} &\multicolumn{1}{c}{30} &\multicolumn{1}{c}{50} &\multicolumn{1}{c}{100} \\
\hline
20  & 0.233 &-0.251 &-0.282 & 0.004  &&0.077 &0.063 &0.048 &0.033 \\
30  & 0.124 & 0.095 & 0.055 & 0.031  &&0.062 &0.050 &0.038 &0.027\\
50  & 0.197 &-0.043 & 0.010 & 0.039  &&0.049 &0.038 &0.030 &0.022 \\
100 & 0.123 & 0.072 &-0.034 &-0.029  &&0.033 &0.027 &0.021 &0.014 \\
\hline
\end{tabular}}
\end{table}

\begin{table}[!ht]  		
  \centering
  \caption{Median of Bias and RMSE of the nonparametric estimator in (\ref{eqII6})}\label{tabIV2}
  \resizebox{12cm}{!}{
  \begin{tabular}{lrrrrrrrrr}
  \hline
&\multicolumn{4}{c}{Bias} &&\multicolumn{4}{c}{RMSE}\\
\hline
N/T   &\multicolumn{1}{c}{20} &\multicolumn{1}{c}{30} &\multicolumn{1}{c}{50} &\multicolumn{1}{c}{100}  &&\multicolumn{1}{c}{20} &\multicolumn{1}{c}{30} &\multicolumn{1}{c}{50} &\multicolumn{1}{c}{100} \\
\hline
20  &-0.102 &-0.090 &-0.072 &-0.054 &&0.235 &0.201 &0.170 &0.132\\
30  &-0.087 &-0.075 &-0.057 &-0.045 &&0.202 &0.174 &0.148 &0.115\\
50  &-0.070 &-0.059 &-0.044 &-0.034 &&0.169 &0.148 &0.121 &0.097\\
100 &-0.053 &-0.046 &-0.035 &-0.023 &&0.134 &0.116 &0.098 &0.079\\
\hline
\end{tabular}}
\end{table}

Table \ref{tabIV1} tells us that the proposed estimator for $\beta_1$ seems to perform quite well in finite samples. As was expected from the results in Theorem \ref{theo1}, the RMSEs of the estimator decrease as both $N$ and $T$ grow. Similar behavior is observed in Table \ref{tabIV2} for the nonparametric estimator proposed for $g(w_{it})$ corroborating the results in Theorem \ref{theo2}.

To assess size and power of our test consider DGP (\ref{eqIV1}) with $m_0(\lambda_i,f_t)$ under $H_0$ and with $m_0^*(\lambda_i,f_t)=m_0(\lambda_i,f_t)+\Delta_N(4x_{it}^2+z_t^3-3w_{it})$ under $H_1$, for $0\leq\Delta_N\leq1$. This specification of $m_0^*(\lambda_i,f_t)$ enables us to evaluate the testing power over different values of $\Delta_N$ when it departs from the null ($\Delta_N=0$), by increasing $\Delta_N$ to $1$ at step length of $0.05$.

\begin{table}[htb]
  \centering
  \caption{Proportion of rejections under $H_0$ based on bootstrap p-values.}\label{tabIV5}
  \resizebox{6cm}{!}{
\begin{tabular}{lcccc}
  \hline
  N/T &\multicolumn{1}{c}{20} &\multicolumn{1}{c}{30} &\multicolumn{1}{c}{50} &\multicolumn{1}{c}{100}\\
  \hline
  20  &0.061 &0.054 &0.061 &0.059\\
  30  &0.054 &0.067 &0.076 &0.051\\
  50  &0.053 &0.079 &0.064 &0.071\\
  100 &0.072 &0.067 &0.046 &0.057\\
\hline
\end{tabular}}
\end{table}

\begin{figure}
\centering
{\includegraphics[width=0.7\textwidth]{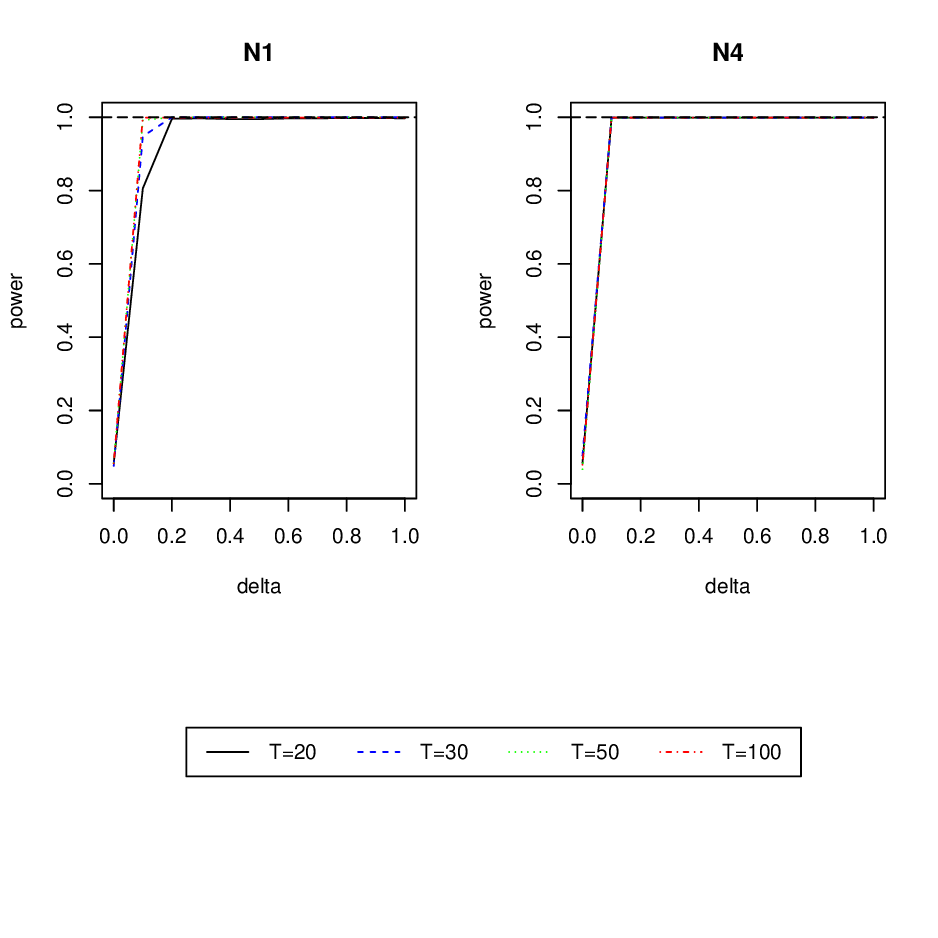}}
\caption{\label{fig:test_power} Power curves under $H_1$. }
\end{figure}

Table \ref{tabIV5} shows the obtained size for the proposed test at $1\%$, $5\%$, and $10\%$ significance levels, where critical values were calculated by wild bootstrap. Also, in Figure \ref{fig:test_power} we plot the power functions of the test statistic against $\Delta_{N}$ for sample sizes $N1=20$ and $N4=100$ at $5\%$ significance level. Note that the results for the other percentile values (i.e., $10\%$ and $1\%$) and the power functions for the different values of $N$ are collected in the Supplement.

Looking at the results in Table \ref{tabIV5}, we see that our test statistic $V_{NT}$ holds reasonably well for all sample sizes considered and for all percentile values of the null distribution of our test statistic. Also, Figure \ref{fig:test_power} shows an impressive power of our test already for small and moderate samples, increasing rapidly with $\Delta_N$ even for the smallest value of $N$.

\section{Estimating Effects of the EU-ETS} \label{secV} \setcounter{equation}{0}

We employ our methods to assess the effect of the European Union Emissions Trading System (EU-ETS) to decarbonize the economy and mitigate environmental degradation. As it is well-known, climate change is today's greatest environmental challenge and social concern. 
In recent years, the European Commission has been one of the most decisive institutions in leading the global energy transition, supporting the achievement of a low-carbon economy through targets and regulatory policies. The implementation of the EU-ETS in $2005$ has become the cornerstone of the European Union's strategy to decarbonize the economy and mitigate climate change \citep{BORGHESI_FRANCO_MARIN_2020}. Nevertheless, despite the efforts made recently towards decarbonization, the burning of fossil fuels (carbon, oil, and gas) related to countries' economic activity continues to increase CO$_2$ emissions.

The relationship between economic development and environmental quality is an issue that has long puzzled economists, and there is a long tradition of employing the Environmental Kuznets Curve (EKC) to shed light on this issue. The EKC is based on the concept proposed by \citet{KUZNETS:1955} which posits an inverted U-shape relationship between income and environmental degradation \citep{GROSSMAN_KRUEGER:1993,GROSSMAN_KRUEGER:1995}. In other words, the EKC postulates that environmental degradation rises with income during the initial phases of economic growth when income is relatively low since industry development causes great damage to the environment's quality. However, after passing a certain income threshold, this relationship reverses and becomes a negative one since higher levels of development are associated with a change in the economic structure in favor of industry and services that are more efficient and environmentally friendly. Since the pioneering work of \citet{GROSSMAN_KRUEGER:1993} several studies have tried to corroborate this EKC hypothesis, but so far there is no consensus on this relationship (see \citet{DINDA:2004}, \citet{GALEOTTI_LANZA_PAULI:2006}, or \citet{KAIKA_ZERVAS:2013a,KAIKA_ZERVAS:2013b}, among others). This lack of consensus may be due to several misspecification errors related to the omission of relevant variables that can lead to inconsistent estimators and misleading inferences.

In this context, we propose to assess the effect of EU-ETS on CO$_2$ emissions through an extension of the conventional EKC specification based on the Stochastic Impacts by Regression on Population, Affluence and Technology (STIRPAT) model for evaluating environmental change that tries to overcome the above shortcomings \citep{DIETZ_ROSA:1997}. On the one hand, we augment the model with a common stochastic covariate, $z_t$, to control for the aggregate effect of EU-ETS carbon pricing on CO$_2$. On the other hand, we propose to control for relevant omitted variables by allowing interactive fixed effects (i.e., $m_0\left(\lambda_i,f_t\right)$, where $f_t$ denotes a vector of unobserved common factors and $\lambda_i$ are the corresponding factor loadings that are allowed to be heterogeneous among countries $i$). Hence, the resulting panel data model to consider would be
\begin{eqnarray}\label{eqV1}
\nonumber ln(CO_{2it})&=&\beta_0+\beta_1ln(gdp_{it})+\beta_2ln(gdp_{it})^2+\beta_3ln(enit_{it})+\beta_4ln(pop_{it})\\
&+&\theta_1ln(z_t)+m_0\left(\lambda_i,f_t\right)+u_{it},
\end{eqnarray}
where $CO_{2it}$ denotes the emissions of fossil CO$_2$ of country $i$ at time $t$, $gdp$ is the gross domestic product,
$enit$ denotes technology which is proxied by energy intensity to capture technology's damaging effect on the environment, $pop$ denotes the population size, $z_t$ is the price of the carbon emissions set by the EU-ETS to all the country's members, and $u_{it}$ captures the innovation. Note that all the variables in (\ref{eqV1}) are expressed in natural logarithms, so the estimated coefficients are all interpreted as elasticities.

The data used for this study covers EU27 countries plus the UK
 and Norway over the period $2005$-$2021$. The variables are derived from official sources and the specific definitions are detailed in Section S.2 of the Supplement.  
As control variables $w_{it}$ we use the natural logarithm of the total
stock of debt liabilities as share of DGP that is collected by the International Monetary Fund and the age dependency ratio from the World Bank database.

The $\beta$ estimates  are reported in Table \ref{tabV2}. We also provide estimates that we obtained using the CCE approach of \citet{PESARAN:2006} and  naive estimates when unobserved factors were simply ignored. This was done to assess the potential misspecification problems related to the functional form of the interactive fixed effects. Estimates using the approach of \citet{BAI:2009} are not provided since the policy variable that we are evaluating in this paper is individual-invariant, and the proposed bootstrap in \citet{BAI:2009} does not converge.

\begin{table}[htb]
  \centering
  \caption{Parametric estimates}\label{tabV2}
  \resizebox{9cm}{!}{
  \begin{tabular}{llll}
  \hline
  &$\widehat{\beta}$ &$\widehat{\beta}^*$ &$\widehat{\beta}_{CCE}$\\
  \hline
  constant       & \mbox{}      &3.285$^{***}$  &-0.531\\
                 &  \mbox{}      &(3.43e-07)     &(0.947)\\
  gdp            &0.878$^{***}$  &0.512$^{**}$   &-0.106\\
                 &(9.21e-13)     &(0.001)        &(0.088)\\
  gdp$^2$        &-0.437$^{***}$ &-0.174         &1.926$^{**}$\\
                 &(2.51e-04)     &(0.277)        &(0.004)\\
  enit           &0.051$^{***}$  &0.083$^{***}$  &0.172$^{***}$\\
                 &(5.07e-09)     &(1.38e-12)     &(2.2e-16)\\
  pop            &0.644$^{***}$  &0.657$^{***}$  &-1.217$^{***}$\\
                 &(2e-16)        &(2e-16)        &(3.81e-05)\\
  EU ETS         &3.24e-04       &-7.9e-03       &4.05e-05\\
                 &(0.769)        &(0.470)        &(0.975)\\
  EKC Holds                 &Yes            &No            &No\\
  Turning points            &1.212 &1.046 &2.782\\
  (millions of 2015 $\$$US) & & &\\
  \hline
  \end{tabular}}
  \begin{minipage}{10cm}
  \vspace{0.1cm}
  \scriptsize \textbf{Note:} $\widehat{\beta}$ is our estimator, $\widehat{\beta}^*$ is the naive estimator, and $\widehat{\beta}_{CCE}$ is the pooled CCE estimator of \citet{PESARAN:2006}. P-values are in parenthesis. $^{***}$ and $^{*}$ indicates statistical significance at the $1\%$ and $10\%$, respectively. Test statistic for our model is $-0.484$ with p-value $0.740$.
  \end{minipage}
\end{table}

The results in the first column of Table \ref{tabV2} corroborate most of the postulates of the empirical literature (see \citet{DINDA:2004}, \citet{LANTZ_FENG:2006}, \citet{CHURCHILL_INEKWE_IVANOVSKI_SMYTH:2018}, \citet{WANG_ZHANG_LI:2023}, among others). On the one hand, the EKC hypothesis (i.e., a positive coefficient of $gdp$ and a negative one for $gdp^2$) for the income-pollution relationship is obtained, with the turning point at $gdp^* = \exp(-\beta_1/2\beta_2)$. On the other hand, the significant positive effect of population and energy intensity on environmental degradation is proven.

Finally, the results in Table \ref{tabV2} enable us to endorse the inability of pre-established standard functional forms for interactive effects to capture certain non-linearities of the effects of unobserved common factors on individuals. More precisely, looking at the results in column (ii) we can see that the EKC hypothesis does not hold when we completely ignore the unobserved common factors, and the $gdp^2$ is not statistically relevant. In addition, when we follow \citet{PESARAN:2006}'s approach to control for the unobserved common factors, the effects of $gdp$ and $gdp^2$ are not the expected ones and are not statistically significant. Moreover, p-value of the test statistic related to the specification test that is proposed to check the crucial modeling assumption of the common factors is $0.740$. Therefore, the $H_0$ is not rejected supporting thereby our modeling proposal.

\section{Conclusion} \label{secVI}

In this paper, identification, inference, and validation of a linear panel data model is considered that allows for an unspecified factor structure. This avoids the strong parametric restrictions that usually appear in the error term of traditional panel data models. The introduction of a conditional mean independence restriction between factors, factor loadings, and covariates converts the model into a partial linear model.  A specification test to verify the crucial identification assumption made is introduced. For the parameters that attract most of the attention, consistent estimators that are asymptotically normal at the optimal rate are derived. Th specification test relies on combining the methodology of conditional moment tests and nonparametric estimation techniques. Using degenerate and nondegenerate theories of U-statistics we can show its convergence, asymptotic distribution under the null, and its divergence under the alternative at a rate arbitrarily close to $\sqrt{NT}$. The good performance of our estimators and test is confirmed by Monte Carlo experiments. Finally, the proposed approach is used to assess the effect of the EU ETS on CO2 emissions and the economic development of EU countries. The findings exhibit the practical relevance of our approach to avoid misleading conclusions, and they show the inability of existing popular methods to control well for the confounding effects of unobserved common factors.

\bibliographystyle{Chicago}
    \bibliography{NPCSD_bib}

    \section*{SUPPORTING INFORMATION}
Additional supporting information may be found in the Supplement of this article at the publisher's website. Section S1 includes the extended Mont Carlo experiment aiming to analyze the finite-sample performance of the proposed estimator and test statistic. Section S2 describes the data set used for the empirical application. Section S3 presents and proves some lemmas needed to show the main results. 


    \section*{Appendix A: Proofs of Results}\label{appendixA_main}
    \renewcommand{\theequation}{A.\arabic{equation}}
    \renewcommand{\thesection}{A}
    \setcounter{equation}{0}


\textbf{Proof of Theorem \ref{theo1}:} Write (\ref{eqI1a}) in vectorial form and plug the resulting expression in (\ref{eqII5}), the expression to analyze the asymptotic properties of $\widehat{\beta}$ becomes
\begin{eqnarray}\label{eqA1}
\sqrt{NT}\left(
\widehat{\beta}-\beta\right)=[(NT)^{-1}\widetilde{X}^{\prime}\widetilde{X}]^{-1}\frac{1}{\sqrt{NT}}\widetilde{X}^{\prime}(I_{NT}-S)\left[m_0(\lambda,F)+U\right].\qquad
\end{eqnarray}
Using the results in Lemma S3.1 and the Law of Large Numbers (LLN), $(NT)^{-1}\widetilde{X}^{\prime}\widetilde{X}\overset{p}{\to}\Omega_x$ and the numerator term to be analyzed can be written as
\begin{eqnarray}\label{eqA2}
(NT)^{-1/2}\widetilde{X}^{\prime}(I_{NT}-S)\left[m_0(\lambda,F)+U\right]= \frac{1}{\sqrt{NT}}\sum_{i=1}^N\sum_{t=1}^T\{x_{it}-B_x(w_{it})\}\epsilon_{it}+o_p(1),\qquad
\end{eqnarray}
where $\epsilon_{it}=m_0\left(\lambda_i,f_t\right)-g(w_{it})+u_{it}$. Under Assumptions \ref{asumA1}-\ref{asumA3}, and using the properties of the law of iterated expectations (LIE), it is easy to show that
\begin{align*}
&E[(NT)^{-1/2} \sum_{it}\{x_{it}-B_x(w_{it})\}\epsilon_{it}] = 0
\qquad \text{ and } \\
&Var[(NT)^{-1/2}\sum_{it}\{x_{it}-B_x(w_{it})\}\epsilon_{it}]=\Sigma_{1,\epsilon}.
\end{align*}
Therefore, using the above results we can conclude that
\begin{eqnarray}\label{eqA3}
Var\left[\sqrt{NT}\left(
\widehat{\beta}-\beta\right)\right]=\Omega_x^{-1}\Sigma_{1,\epsilon}\Omega_x^{-1}\{1+o_p(1)\}.
\end{eqnarray}
To finish the proof of the theorem we have to consider the asymptotic distribution of the stochastic term of (\ref{eqA1}). When $T$ is fixed, we use the Lindeberg-Feller Central Limit Theorem (CLT) for $i.i.d.$ random variables so we can conclude
\begin{eqnarray}\label{eqA4}
\sqrt{NT}\left(
\widehat{\beta}-\beta\right)\overset{d}{\to}\mathcal{N}(0,\Omega_x^{-1}\Sigma_{1,\epsilon}\Omega_x^{-1}).
\end{eqnarray}
On its part, considering the case in which $T\rightarrow\infty$, we employ the Cram\'er-Wold device. Let $\varpi_{At}=N^{-1/2}\sum_{i=1}^N\imath^{\prime}\ddot{X}_{i\cdot}\epsilon_{i\cdot}$ for any unit vector in $\imath\in\mathbb{R}^{d_x}$, where $\ddot{X}_{i\cdot}$ is a $T\times d_x$ matrix whose $t$-th element is such as $\ddot{x}_{it}=x_{it}-B_x(w_{it})$, and let $\epsilon_{i\cdot}\equiv(\epsilon_{i1},\ldots,\epsilon_{iT})$ be a $T\times 1$ vector. Then we can write
\begin{eqnarray*}
(NT)^{-1/2}\sum_{i=1}^N\sum_{t=1}^T\{x_{it}-B_x(w_{it})\}\epsilon_{it}=T^{-1/2}\sum_{t=1}^T\varpi_{At}.
\end{eqnarray*}
By Assumption \ref{asumB1}, $\{x_{it}-B_x(w_{it})\}\epsilon_{it}$ is a $\mathbb{R}^{d_x}$-stationary $\alpha$-mixing sequence, and by Assumptions \ref{asumA3}-\ref{asumA4} and \ref{asumB1}-\ref{asumB3}, the Theorem $1.5$ of \citet{BOSQ:1998}, pag.$34$, gives
\begin{eqnarray*}
Var\left(\frac{1}{\sqrt{T}}\sum_{t=1}^T\varpi_{At}\right)
=\plim_{T\rightarrow\infty}\frac{1}{T}\imath^{\prime}E[\ddot{X}_{i\cdot}\epsilon_{i\cdot}\epsilon_{i\cdot}^{\prime}\ddot{X}_{i\cdot}^{\top}]\imath+o_p(1) \ .
\end{eqnarray*}
A direct application of the CLT for strongly mixing processes (see Theorem $1.7$ in \citet{BOSQ:1998}, pag. $36$) gives
\begin{eqnarray*}
\frac{1}{\sqrt{T}}\sum_{t=1}^T\varpi_{At}\overset{d}{\to}\mathcal{N}(0,\imath^{\top}\Sigma_{2,\epsilon}\imath)\quad\textrm{as}\quad N,T\rightarrow\infty.
\end{eqnarray*}
Finally, combining the above results with the Slutsky theorem concludes the proof.
 \hfill$\square$\\

\textbf{Proof of Theorem \ref{theo2}}
Writing (\ref{eqI1a}) in vectorial form, and replacing the resulting expression in (\ref{eqII6}), the Taylor expansion gives
\begin{eqnarray}\label{eqA5}
\nonumber&&\sqrt{NTh_w^{d_w}}\{\widehat{g}(w;h_w)-g(w)\}\\
\nonumber&&-\sqrt{NTh_w^{d_w}}(1,0_{d_w})[\mathcal{W}^{\prime}K_w\mathcal{W}]\mathcal{W}^{\prime}K_w[Q_w(w)+\mathcal{R}_w(w)-X(\widehat{\beta}-\beta)]\\
&&=\sqrt{NTh_w^{d_w}}(1,0_{d_w})[\mathcal{W}^{\prime}K_w\mathcal{W}]\mathcal{W}^{\prime}K_w\epsilon,
\end{eqnarray}
where $Q_w(w)=\frac{1}{2}[(w_{11}-w)^{\prime}\mathcal{H}_g(w)(w_{11}-w),\ldots,(w_{NT}-w)^{\prime}\mathcal{H}_m(w)(w_{NT}-w)]^{\prime}$, $\mathcal{R}_w(w)$ is the residual term of the Taylor expansion, and $\epsilon\equiv(\epsilon_{11},\ldots,\epsilon_{NT})$ is a $NT\times 1$ vector. In the following, we prove that the terms on the left-hand side of (\ref{eqA5}) contribute to the asymptotic bias, whereas the term on the right-hand side is asymptotically normal.

Focusing on the behavior of the inverse term, and following a similar proof scheme as in \citet{SOBERON_MUSOLESI_RODRIGUEZ-POO:2023} or \citet{CAI_LI:2008}, under Assumptions \ref{asumA1}-\ref{asumA4} it can be  shown
\begin{eqnarray}\label{eqA6}
(NT)^{-1}\mathcal{W}^{\prime}K_w\mathcal{W}=\left(\begin{array}{cc}
\rho_w(w)+o_p(1) &h_w^2\mu_2^{d_w}D_{\rho}(w)+o_p(h_w^2)\\
h_w^2\mu_2^{d_w}D_{\rho}(w)+o_p(h_w^2) &h_w^2\mu_2^p\rho_w(w)I_{d_w}+o_p(h_w^2)
\end{array}\right), 
\end{eqnarray}
with $I_{d_w}$ the $d_w\times d_w$ identity matrix, and $D_{\rho}(w)$ the vector of first-order derivatives of $\rho_w$.

Similarly, following the proof for Lemma B2 in \citet{SOBERON_MUSOLESI_RODRIGUEZ-POO:2023}, and with $\sqrt{NT}$-consistency of $\widehat{\beta}$, using the Cram\'er-Wold device it can be shown that the asymptotic bias of $\widehat{g}(w;h_w)$ is of the form
\begin{eqnarray}\label{eqA7}
\nonumber&&\sqrt{NTh_w^{d_w}}(1,0_{d_w})[\mathcal{W}^{\prime}K_w\mathcal{W}]\mathcal{W}^{\prime}K_w[Q_w(w)+\mathcal{R}_w(w)-X(\widehat{\beta}-\beta)]\\
&&=\frac{\mu_2^{d_w}}{2}tr\{h_w^2\mathcal{H}_g(w)\}+O_p(h_w^2)+o_p\left(\frac{1}{\sqrt{NTh_w^{d_w}}}\right)  .
\end{eqnarray}
Considering now the variance term of $\widehat{g}(w;h_w)$, under Assumptions \ref{asumA1}-\ref{asumA3}, and using the LIE, it can be proved $E[(NT)^{-1}\mathcal{W}^{\prime}K_w\epsilon]=0$ and
\begin{eqnarray}\label{eqA8}
\nonumber &&NTh_w^{d_w}Var[(NT)^{-1}\mathcal{W}^{\prime}K_w\epsilon]\\
&&=\left(\begin{array}{cc}
\nu_0^{d_w}\rho_w(w)\frac{1}{T}\sum_{tt'}E(\epsilon_{it}\epsilon_{it'}|w_{it},w_{it'}) &O_p(h_w^{d_w})\\
O_p(h_w^{d_w}) &h_w^2\nu_2^{d_w}\rho_w(w)\frac{1}{T}\sum_{tt'}E(\epsilon_{it}\epsilon_{it'}|w_{it},w_{it'})    \end{array}\right).  \qquad\quad  
\end{eqnarray}
Using (\ref{eqA6}) and (\ref{eqA8}), the Slutsky theorem enables us to conclude that the variance term of $\widehat{g}(w;h_w)$ when $T$ is fixed and $N\rightarrow\infty$, is
\begin{eqnarray}\label{eqA9}
Var\{\widehat{g}(w;h_w)\} = \frac{\nu^{d_w}\Psi_{1,\epsilon}}{NT^2h_w^{d_w}\rho_w(w)}\{1+o_p(1)\} ,
\end{eqnarray}
and the corresponding variance for $N$ and $T$ large is such that
\begin{eqnarray}\label{eqA10}
Var\{\widehat{g}(w;h_w)\} = \frac{\nu^{d_w}\Psi_{2,\epsilon}}{NTh_w^{d_w}\rho_w(w)}\{1+o_p(1)\}  .
\end{eqnarray}
To conclude the asymptotic properties of $\widehat{g}_w(w;h_w)$ we need to establish  asymptotic normality for $\sqrt{NTh_w^{d_w}}(NT)^{-1}\mathcal{W}^{\prime}K_w\epsilon$. To this end, when $T$ is fixed and $N\rightarrow\infty$ we employ the Minkowski inequality to check that the Lyapunov condition holds, and the first part of Theorem \ref{theo2} is proved. Finally, when $N$ and $T$ are large, then we follow a similar proof scheme as in \citet{SOBERON_MUSOLESI_RODRIGUEZ-POO:2023}. Therefore, by employing the Cram\'er-Wold device and the Doob's small-black and large-block technique for $\alpha$-mixing processes, we can conclude that the second part of Theorem \ref{theo2} is proven too.
 \hfill$\square$\\

\textbf{Proof of Theorem \ref{theo5}} \label{secA}
The test statistic (\ref{eqvn}) can be rewritten as
\begin{equation}\label{eqa1}
    V_{NT} = \frac{1}{N^2T^2h^d}\sum^N_{i=1}\sum^N_{ j=1, j\ne i }\widehat{\epsilon}_{i\cdot}^{\prime} \mathbf{K}_{ij}\widehat{\epsilon}_{j\cdot},
\end{equation}
where, for $i=1,\cdots,N$ and $j=1,\cdots,N$, $\widehat{\epsilon}_{i\cdot}= \left(\widehat{\epsilon}_{i1},\cdots,\widehat{\epsilon}_{iT} \right)^{\prime}$ and
\begin{equation*}
\mathbf{K}_{ij} = \left(\begin{array}{ccc}
    K_{i1,j1} & \cdots & K_{i1,jT} \\
   \vdots  &  \cdots  &  \vdots \\
    K_{iT,j1} & \cdots & K_{iT,jT}
\end{array}\right).
\end{equation*}
Then, if we substitute (\ref{eqI1b}) into $\widehat{\epsilon}_{i\cdot} = y_{i} - X_{i\cdot}\widehat{\beta} - \widehat{g}(w_{i\cdot};h_w)$ we obtain $\widehat{\epsilon}_{i\cdot} = \epsilon_{i\cdot}  - \left\{ \widehat{g}(w_{i\cdot};h_w) - g(w_{i\cdot}) \right\}
- X_i(\widehat{\beta}-\beta)$, where $\widehat{g}(w_{i\cdot};h_w) = \left(\widehat{g}(w_{i1};h_w),\cdots,\widehat{g}(w_{iT};h_w)\right)^{\prime}$,  
$g(w_{i\cdot}) = \left(g(w_{i1}),\cdots,g(w_{iT})\right)^{\prime}$,
$\epsilon_{i\cdot} = \left(\epsilon_{i1},\cdots,\epsilon_{iT}\right)^{\prime}$, and $X_{i\cdot}$ is a $T\times d_x$ matrix, for $i=1,\cdots,N$.

The statistic (\ref{eqvn}) can be decomposed into the following six parts:
\begin{eqnarray*}
    V_{NT} =  \frac{1}{N^2T^2h^d}\sum^N_{i=1}\sum_{j\neq i}^N\widehat{\epsilon}_{i\cdot}^{\prime} \mathbf{K}_{ij}\widehat{\epsilon}_{j\cdot}    &=&  
    \frac{1}{N^2T^2h^d}\sum^N_{i=1}\sum_{j\neq i}^N\epsilon_{i\cdot}^{\prime} \mathbf{K}_{ij}\epsilon_{j\cdot} \\
    & + & (\widehat\beta-\beta)^{\prime}\left\{\frac{1}{N^2T^2h^d}\sum^N_{i=1}\sum_{j\neq i}^N X^{\prime}_i\mathbf{K}_{ij}X_j\right\}(\widehat\beta-\beta) \\
    & + & \frac{1}{N^2T^2h^d}\sum^N_{i=1}\sum_{j\neq i}^N \left(\widehat{g}(w_{i\cdot};h_w)-g(w_{i\cdot})\right)^{\prime}\mathbf{K}_{ij}\left(\widehat{g}(w_{j\cdot};h_w)-g(w_{j\cdot})\right) \\
    & - & \frac{2}{N^2T^2h^d}\sum^N_{i=1}\sum_{j\neq i}^N \epsilon^{\prime}_{i\cdot} \mathbf{K}_{ij}X_{j\cdot}(\widehat\beta-\beta) \\
    & - & \frac{2}{N^2T^2h^d}\sum^N_{i=1}\sum_{j\neq i}^N\epsilon^{\prime}_{i\cdot} \mathbf{K}_{ij}\left(\widehat{g}(w_{j\cdot};h_w)-g(w_{j\cdot})\right) \\
    & - & 2(\widehat\beta-\beta)^{\prime}\left\{\frac{1}{N^2T^2h^d}\sum^N_{i=1}\sum_{j\neq i}^NX^{\prime}_i \mathbf{K}_{i,j} \left(\widehat{g}(w_{j\cdot};h_w)-g(w_{j\cdot})\right)\right\} .
    \end{eqnarray*}
Hence, $V_{NT}$ can be rewritten as
\begin{equation}
    V_{NT} \equiv V_{1,NT} + V_{2,NT} + V_{3,NT} - 2V_{4,NT} - 2V_{5,NT} - 2V_{6,NT}.
\end{equation}
Under the null, we will show that $NTh^{d/2}V_{1,NT}$, conditional on $f_1,\cdots,f_T$, is normally distributed, and $NTh^{d/2}V_{l,NT} = o_p(1)$ for $l=2,\cdots,6$. By applying Lemmas 3.2 and 3.3 from the Supplement we obtain the desired result.
\hfill$\square$\\

\textbf{Proof of Theorem \ref{theo6}} \label{secB}

Following the same lines as in the proof of Theorem \ref{theo5}, 
\begin{eqnarray*}
    V_{NT} & = & \frac{1}{N^2T^2h^d}\sum^N_{i=1}\sum_{j\neq i}^N\widehat{\epsilon}_{i\cdot}^{\prime} \mathbf{K}_{ij}\widehat{\epsilon}_{j\cdot} \\
    & = & \frac{1}{N^2T^2h^d}\sum^N_{i=1}\sum_{j\neq i}^N\epsilon_{i\cdot}^{\prime} \mathbf{K}_{ij}\epsilon_{j\cdot} + o_p(1/NTh^{d/2})
    \equiv  
    V_{1,NT} + o_p(1/NTh^{d/2})  .
\end{eqnarray*}
Recall that
\begin{eqnarray*}
TV_{1,NT} & = & \frac{N-1}{N}\left\{\frac{2}{N(N-1)}\sum^N_{i=1}\sum_{i<j}^N \frac{1}{Th^d} \epsilon_{i\cdot}^{\prime} \mathbf{K}_{ij}\epsilon_{j\cdot}\right\}   \equiv  
\frac{N-1}{N}\mathbb{U}_N  ,
\end{eqnarray*}
where
\begin{equation}
    \mathbb{U}_N = \frac{2}{N(N-1)} \sum^N_{i=1}\sum_{i<j}^N\mathbb{H}_{NT}\left(\zeta_i,\zeta_j\right),
\end{equation}
for $\zeta_i = \left(\chi_i,\epsilon_i\right)$ for $i=1,\cdots,N$, and
\begin{eqnarray}
    \mathbb{H}_{NT}\left(\zeta_i,\zeta_j\right) & = & \frac{1}{Th^d} \epsilon_{i\cdot}^{\prime} \mathbf{K}_{ij}\epsilon_{j\cdot}      = 
    \frac{1}{Th^d} \sum^T_{t=1}\sum^T_{s=1}K_{it,js}\epsilon_{it}\epsilon_{js}.
\end{eqnarray}
Like in the proof of Theorem \ref{theo5}, we execute the analysis conditional on the factors $\mathcal{D}_f=\left\{f_t: t=1,\cdots,T\right\}$. By doing so, we get independence among different cross-sectional units.

From the proof of Theorem \ref{theo5} we have that
\[
E_{\mathcal{D}_f}\left[\left\|\mathbb{H}_{NT}\left(\zeta_i,\zeta_j\right)\right\|^2\right] = O\left(\frac{1}{h^d}\right) = O\left(\frac{N}{Nh^d}\right) = o(N) ,
\]
because $Nh^d\rightarrow \infty$. Then, applying Lemma 3.1 of \citet{POWELL_STOCK_STOKER:1989} we obtain
\[
\mathbb{U}_N = E_{\mathcal{D}_f}\left[\mathbb{H}_{NT}\left(\zeta_i,\zeta_j\right)\right] + o_p(1)  .
\]
Since
\begin{eqnarray*}
E\left[\left.\mathbb{H}_{NT}\left(\zeta_i,\zeta_j\right)\right|\zeta_i\right]
& = & \frac{1}{Th^d}\sum^T_{t=1}\sum^T_{s=1}\epsilon_{it}E\left[K\left(\frac{\chi_{it}-\chi_{js}}{h_1}\right)E\left(\left.\epsilon_{js}\right|\chi_{js}\right)\right] \\
& = & \sum^T_{t=1}\epsilon_{it}\left\{E\left(\left.m_0\left(\lambda_i,f_t\right)\right|\chi_{it}\right)-g\left(w_{it}\right)\right\}\rho_{\chi}\left(\chi_{it}\right),
\end{eqnarray*}
one has
\begin{eqnarray*}
E_{\mathcal{D}_f}\left[\mathbb{H}_{NT}\left(\zeta_i,\zeta_j\right)\right] & = &
E_{\mathcal{D}_f}\left[E\left[\left.\mathbb{H}_{NT}\left(\zeta_i,\zeta_j\right)\right|\zeta_i\right]\right] \\
& = & \sum^T_{t=1}E_{\mathcal{D}_f}\left[\left\{E\left(\left.m_0\left(\lambda_i,f_t\right)\right|\chi_{it}\right)-g\left(w_{it}\right)\right\}^2\rho_{\chi}\left(\chi_{it}\right)\right]  .
\end{eqnarray*}
Therefore, under $H_1$,
\[
V_{1,NT} = \frac{1}{T}\sum^T_{t=1}E_{\mathcal{D}_f}\left[\left\{E\left(\left.m_0\left(\lambda_i,f_t\right)\right|\chi_{it}\right)-g\left(w_{it}\right)\right\}^2\rho_{\chi}\left(\chi_{it}\right)\right] + o_p(1).
\]
\hfill$\square$\\

\end{document}